\documentclass[aps,twocolumn,prb]{revtex4}

\usepackage{graphicx}
\usepackage{dcolumn}
\usepackage{bm}

\begin{document}
\preprint{APS/123-QED}

\title{Quantum criticality near the upper critical field of
Ce$_2$PdIn$_8$}

\author{Y. Tokiwa}
\author{P. Gegenwart}

\affiliation{I. Physikalisches Institut,
Georg-August-Universit\"{a}t G\"{o}ttingen, 37077 G\"{o}ttingen,
Germany}

\author{D. Gnida}
\author{D. Kaczorowski}

\affiliation{Institute of Low Temperature and Structure Research,
Polish Academy of Sciences, PO Box 1410, 50-950 Wroc\l aw, Poland}

\date{\today}

\begin{abstract}
We report low-temperature specific heat measurements in magnetic
fields up to 12 T applied parallel and perpendicular to the
tetragonal c-axis of the heavy fermion superconductor
Ce$_2$PdIn$_8$. In contrast to its quasi-two-dimensional (2D)
relative CeCoIn$_5$, the system displays an almost isotropic upper
critical field. While there is no indication for a FFLO phase in
Ce$_2$PdIn$_8$, the data suggest a smeared weak first-order
superconducting transition close to $H_{c2}\approx 2$~T. The normal
state electronic specific heat coefficient displays logarithmically
divergent behavior, comparable to CeCoIn$_5$ and in agreement with
2D quantum criticality of spin-density-wave type.
\end{abstract}

\maketitle

Unconventional superconductivity (SC) often occurs in systems with
competing phases near a quantum critical point (QCP). Indicative for
quantum criticality is e.g. the observation of non-Fermi liquid
(NFL) behavior in the normal state at $T>T_c$ in zero field or down
to the lowest accessible temperatures at fields above the upper
critical magnetic field $H_{c2}$. Heavy-fermion (HF) superconductors
have sufficiently small $H_{c2}$, allowing for a detailed
thermodynamic study of the normal state properties down to very low
temperatures to investigate NFL behavior. The layered tetragonal
CeCoIn$_5$ with $T_c=2$~K is a prototype heavy-fermion
superconductor with a NFL normal state due to a field-induced QCP
slightly below $H_{c2}$.
\cite{petrovic:jpcm-01,bianchi:prl-03b,paglione:prl-03,Paglione-prl06,Zaum-prl11,Donath-prl08,Ronning-prb06,Singh-prl07}
Its SC properties are also fascinating. The transition at
$H_{c2}(T)$ becomes first order below 1~K due to strong Pauli
limiting~\cite{bianchi:prl-02,tayama:prb-02} and clear phase
transition anomalies at the high-field and low-temperature (HFLT)
corner in the SC phase diagram have been
discovered,\cite{bianchi:prl-03a,radovan:nature-03} which were
thought to be related to a modulated SC
Fulde-Ferrell-Larkin-Ovchinnikov (FFLO)
state.\cite{fulde-ferrell:pr-64,larkin-ovchinnikov:jetp-64} The
origin of the HFLT SC state is under extensive debate, as subsequent
NMR and neutron diffraction experiments have revealed the existence
of an antiferromagnetic (AF) ordering within the HFLT
phase,\cite{Kenzelmann-Science08,Young-PRL07,Koutroulakis-prl10}
while a very recent NMR study suggests spatially distributed normal
quasi-particle regions due to the formation of a FFLO
state.\cite{kumagai-prl11}

The recently discovered HF superconductor ($T_c=0.7$~K)
Ce$_2$PdIn$_8$~(Ref.\cite{Kaczorowski-prl09,Kaczorowski-SSS10}) belongs to the same
class of Ce$_n$TIn$_{3n+2}$ (T; transition metal, $n=1,2$ and
$\infty$) systems,\cite{Thompson-JMMM01} containing a series of
CeIn$_3$ and TIn$_2$ layers stacked along the $c$-axis. While the
$n=1$ systems with T=Co, T=Ir (Ref. \cite{petrovic:epl-01}) and T=Rh
 (Ref. \cite{hegger-prl00}) display anisotropic properties due to the
layered, quasi two-dimensional (2D) structure, cubic CeIn$_3$
($n=\infty$) is a completely isotropic system. The bilayer ($n=2$)
systems are therefore expected to display intermediate behavior due
to the more 3D character of their crystal structure consisting of a
stack of two CeIn$_3$ layers separated by one PdIn$_2$ layer.
Ce$_2$PdIn$_8$ is an ambient-pressure HF superconductor of this
class of materials. Its upper critical field $H_{c2}$ is strongly
Pauli limited with a large Maki parameter of 2.9.\cite{Dong-CM11}
Clean single crystals have become available, with a residual
resistivity of about $2.5\mu\Omega$cm and a large mean free path of
$l$=420\,\AA,\cite{Dong-CM11} exceeding the SC coherence length
$\xi$=82\,\AA. Thus, the system fulfills the necessary conditions
for the formation of the FFLO state and it is interesting to
investigate the high-field and low-temperature part of the SC phase
by a thermodynamic probe. Furthermore, recent low-$T$ electrical
resistivity measurements at fields above the upper critical field
have revealed strong similarities to CeCoIn$_5$, i.e. a quasi-linear
dependence at $H_{c2}$ which turns into Fermi liquid behavior only
at larger fields,\cite{Dong-CM11,tran-prb11} suggesting a
field-induced QCP near $H_{c2}$. Because of these common features,
it is highly interesting to study the low-temperature specific heat
of single crystalline Ce$_2$PdIn$_8$ in magnetic fields applied both
parallel and perpendicular to the tetragonal c-axis.

Single crystals were grown by the self-flux method~\cite{Kaczorowski-prl09} and very thin
high-quality pieces were glued on the sample holder. The basal plane
orientation for the different pieces was not investigated. The
specific heat at temperatures down to 70\,mK and magnetic fields up
to 12\,T was measured in a dilution refrigerator with a SC magnet by
employing the quasiadiabatic heat pulse method.

\begin{figure}[htb]
\includegraphics[width=\linewidth,keepaspectratio]{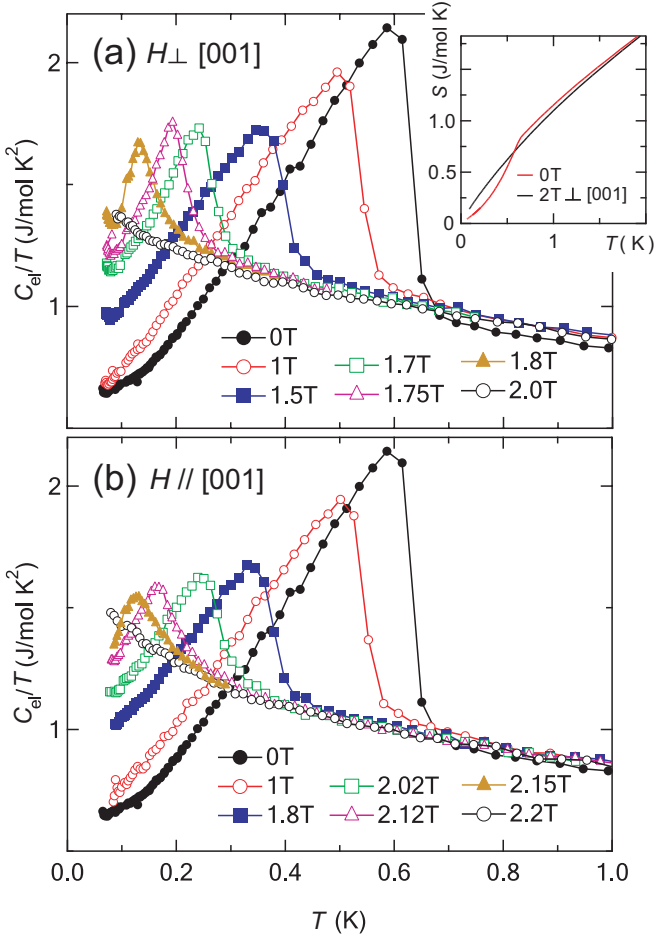}
\caption{(Color online) Electronic specific heat as $C_{el}(T)$/$T$
of Ce$_2$PdIn$_8$ at different magnetic fields perpendicular (a) and
along (b) the c-axis, respectively. The inset in (a) displays the
entropy contributions, obtained from the integral of the electronic
specific heat coefficients at 0 and 2~T, extrapolated down to zero
temperature by assuming a constant (for zero field) and logarithmic
dependence for 2~T (cf. black solid line in Fig.~5a), respectively.}
\end{figure}

Figure~1 shows the specific heat after subtraction of the nuclear
contribution as $C_{el}(T)/T$, at magnetic fields up to $H_{c2}$. The nuclear contribution has been determined from the $\alpha/T^3$ contribution to $C/T$ at lowest temperatures. We have verified, that $\alpha\sim H^2$ as expected for paramagnetic materials. At
zero field, $C_{el}/T$ increases upon cooling to $T_c$=0.68\,K, at
which a typical mean-field type anomaly is found. Within the SC
state, it decreases linearly down to 0.2\,K and saturates at a
residual value of $C_{el}/T=0.65$\,J/mol K$^2$ (0.33\,J/Ce-mol
K$^2$), which is much larger than the respective value of
0.04\,J/mol K$^2$ found for CeCoIn$_5$,~\cite{movshovich:prl-01}
indicating a larger quasi-particle density of states arising from
pair breaking of defects. The specific heat at magnetic fields along
different directions is rather similar and displays only a weak
anisotropy in $H_{c2}=2.0$\,T and 2.2\,T for $H\perp$[001] and
$H\parallel$[001], respectively. At high fields, the SC transition
changes its shape and becomes a rather symmetric peak different to
the expectation for standard superconductors. This may indicate a
smeared first-order transition. Note, that in CeCoIn$_5$ a sharp and
large peak of the specific heat is found at low-temperatures, which
has proven to be due to a first-order
transition.\cite{bianchi:prl-02}

\begin{figure}[htb]
\includegraphics[width=\linewidth,keepaspectratio]{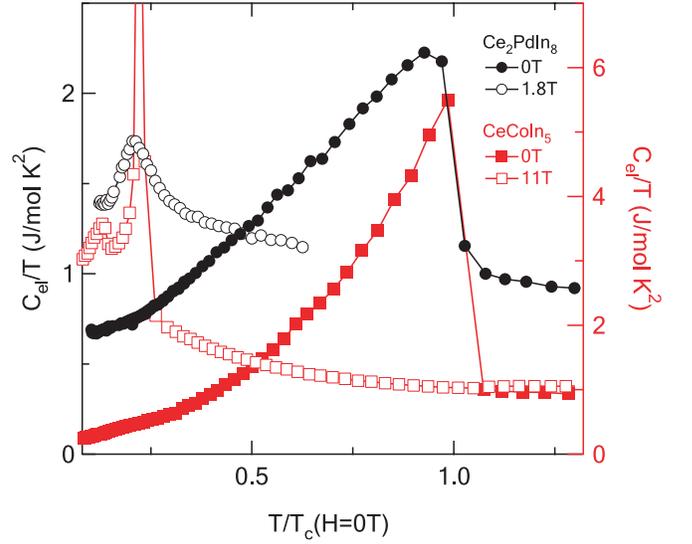}
\caption{(Color online) Comparison on $C_{el}(T)/T$ for
Ce$_2$PdIn$_8$ (circles, $H\perp c$, left axis) and CeCoIn$_5$
(squares, $H\parallel c$, right axis),~\cite{tokiwa-prl08} plotted
against the reduced temperature.}
\end{figure}

In Figure~2, we compare the zero- and high-field SC transition
anomalies of Ce$_2$PdIn$_8$ and CeCoIn$_5$ (Ref.
~\cite{tokiwa-prl08}) on a reduced temperature scale. At $H=1.8$\,T
for the former and 11\,T for the latter system, respectively, $T_c$
is suppressed to 20\% of its zero-field value. The peak at the SC to
normal transition in the latter system is much sharper indicating a
much stronger first-order nature of SC transition in CeCoIn$_5$. The
peak height for CeCoIn$_5$ has been reported to be very sensitive to
impurity scattering. Only 0.22\,\% of Cd-doping suppresses the sharp
peak in specific heat significantly and leads to a very similar C/T
dependence as found for 1.8~T in Ce$_2$PdIn$_8$.\cite{tokiwa-prl08}.
We may thus associate the reduced peak height for Ce$_2$PdIn$_8$ to
the about ten-times shorter electronic mean free path, compared to
undoped CeCoIn$_5$ (4000~\AA).~\cite{Seyfarth-prl08} The HFLT phase
in undoped CeCoIn$_5$ causes an additional jump in $C_{el}(T)/T$ of
about $\sim 0.2$\,J/mol$\cdot$K$^2$ at 11~T, whereas in
Ce$_2$PdIn$_8$ no anomaly could be detected within the measurement
error of $\sim0.02$\,J/mol$\cdot$K$^2$ at 0.1\,K.

\begin{figure}[htb]
\includegraphics[width=\linewidth,keepaspectratio]{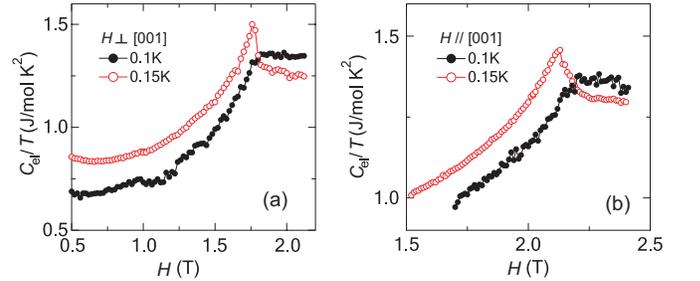}
\caption{(Color online)  Electronic specific heat devided by
temperature, $C_{el}/T$, for Ce$_2$PdIn$_8$ at 0.1 and 0.15\,K as a
function of magnetic field perpendicular (a) and along (b) the
tetragonal $c$-axis, respectively.}
\end{figure}

Measurements of the electronic specific heat coefficient as a
function of magnetic field, displayed in Figure~3, also show no sign
of an anomaly in addition to $H_{c2}$.

\begin{figure}[htb]
\includegraphics[width=\linewidth,keepaspectratio]{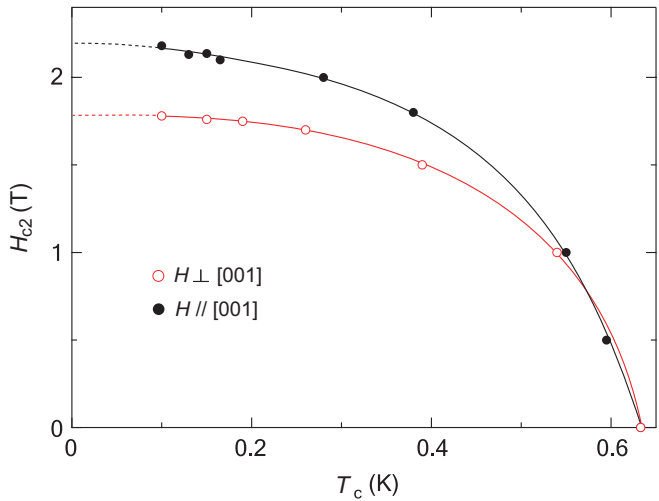}
\caption{(Color online) SC phase diagram of Ce$_2$PdIn$_8$ for two
perpendicular field orientations. The solid lines are guides to the
eyes.}
\end{figure}

Figure~4 shows the phase diagram of Ce$_2$PdIn$_8$ for two
perpendicular magnetic field directions, revealing an almost
isotropic field effect on superconductivity. The strongly convex
shape of $H_{c2}(T)$ indicates a pronounced effect of
Pauli-limiting, dominating over orbital limiting. In CeCoIn$_5$,
$H_{c2}$ is also strongly Pauli limited and furthermore displays a
pronounced anisotropy, which could be related to the anisotropic
normal state spin susceptibility. For Ce$_2$PdIn$_8$, by contrast,
the upper critical field is almost isotropic, in consistent with the much weaker
anisotropy of the reported normal-state susceptibility.~\cite{Uhlirova-Int10}

Next, we turn to the signatures of NFL behavior in the normal state.
As shown in Figure 5, for fields slightly above $H_{c2}$, a
logarithmic divergence of the specific heat coefficient is found
over more than one decade in $T$. The entropy release in zero field
at $T_c=0.68$~K due to the formation of the SC state is almost fully
balanced at $H=2$~T (cf. inset of Fig.~1a), indicating that the same
degrees of freedom are responsible both for the NFL behavior and the
SC state. At $H>H_{c2}$ for both magnetic field directions, Fermi
liquid behavior is gradually recovered, as seen from the evolution
towards a temperature independent specific heat coefficient
$C_{el}/T$. In the insets of Fig. 5, we plot the magnetic field
dependence of the Sommerfeld coefficient, $\gamma(H)$, obtained from
the saturated $C_{el}/T$ values. Upon reducing the field from large
values down to $H_{c2}$, divergent behavior is found in $\gamma(H)$.
This indicates the existence of a field-induced QCP in close
vicinity to the upper critical field of superconductivity, similar
as in CeCoIn$_5$.\cite{bianchi:prl-03b} In order to obtain
information on the position of the QCP, as well as on the nature of
the underlying quantum critical fluctuations, we have carefully
compared the data with the predictions of the Hertz-Millis theory
for an AF QCP in either 2D or 3D. The predicted field dependences
are $\gamma_0+\gamma_h\ln[1/(H-H_c)]$ (2D) and
$\gamma_0-\gamma_h\sqrt{H-H_c}$ (3D), respectively.~\cite{zhu} We
have tried least squares fitting of our data using $\gamma_0$,
$\gamma_h$ and $H_c$ as free parameters. The field-dependence
expected for 3D critical fluctuations appears to be too weak to
explain the experimental $\gamma$ values at sufficiently large
fields. A much better description could be obtained over one decade in $h$=$H$-$H_{c}$, using the
expectation for 2D critical behavior. The obtained quantum
critical fields for the 2D description are $H_c=2.0\pm0.2$\,T and
$1.7\pm0.3$\,T for $H\perp$ and $\parallel$ to [001], respectively.
These critical fields are very close to $H_{c2}$. Note, that recent
electrical resistivity measurements have also suggested a
field-induced QCP very close to $H_{c2}$ in this
system.\cite{Dong-CM11} Furthermore, at $H\approx H_{c2}$ both the
linear temperature dependence of the electrical
resistivity,\cite{Dong-CM11} as well as the logarithmic temperature
dependence of the specific heat coefficient are in agreement with
the theoretical prediction for a 2D AF QCP.

\begin{figure}[htb]
\includegraphics[width=\linewidth,keepaspectratio]{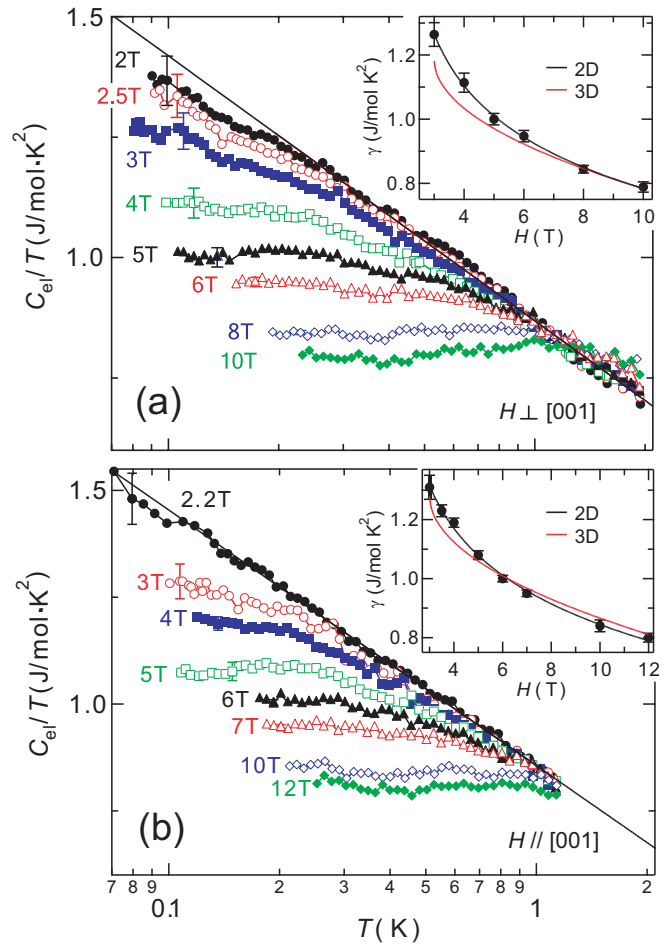}
\caption{(Color online) Electronic specific heat divided by
temperature, $C_{el}(T)/T$, (on a logarithmic scale) of
Ce$_2$PdIn$_8$ at magnetic fields above $H_{c2}$, perpendicular (a)
and along (b) the tetragonal $c$-axis, respectively. The solid lines
represent a $\ln(1/T)$ dependence. The insets display the field
dependence of Sommerfeld coefficient (see text). The colored solid
lines represent the fitted curves using theoretical functions for 2D
and 3D AF QCPs.~\cite{zhu}}
\end{figure}

In summary, we have studied the effect of magnetic fields on the SC
and normal-state behavior in Ce$_2$PdIn$_8$ by measuring the
specific heat at low temperatures and high magnetic fields. Even
though the upper critical field $H_{c2}(T)$ is strongly
Pauli-limited, a phase transition into a HFLT phase as seen in the
relative CeCoIn$_5$ has not been detected. It has been recently
reported that an extremely small amount of impurities destroys (or
smears out) this phase transition in CeCoIn$_5$. For 0.05\,\%
Cd-doped or 0.08\,\% Sn-doped CeCoIn$_5$ it could not be detected by
specific heat measurements any more.\cite{tokiwa-prl08,tokiwa-prb10}
It has also been shown theoretically that a very small amount of
impurities smears out the FFLO phase.\cite{ikeda-prb10} Although
Ce$_2$PdIn$_8$ is a clean limit superconductor, the electronic mean
free path of the investigated single crystal is about ten times
smaller compared to CeCoIn$_5$. This seems reflected in the much
larger residual specific heat coefficient $C_{el}/T$ at zero field
and reduced peak height of the SC transition in high fields. From
the comparison of the respective specific heat data, we find the
investigated Ce$_2$PdIn$_8$ single crystals comparable to 0.22\%
Cd-doped CeCoIn$_5$.\cite{tokiwa-prl08} Thus, a possible HFLT
transition in Ce$_2$PdIn$_8$ could already be destroyed or smeared
out by impurity scattering. Furthermore, we note that the upper
critical field of the latter compound is almost isotropic, whereas
it displays a pronounced anisotropy in CeCoIn$_5$. The observed NFL
normal-state behavior in Ce$_2$PdIn$_8$ appears to be rather similar
as found in CeCoIn$_5$. We have obtained a logarithmic divergence of
$C_{el}(T)/T$ at fields near $H_{c2}$ and a logarithmic field
dependence of the Sommerfeld coefficient, which both are compatible
with 2D AF fluctuations due to a QCP near $H_{c2}$, in close
similarity to CeCoIn$_5$.

This work has been supported by the German Science Foundation
through FOR 960 (Quantum phase transitions).

\bibliography{hf}

\end{document}